
%
%
\magnification=\magstep1
\font\tenss=phvr
\font\sevenss=phvr at 7pt
\font\fivess=phvr at 5pt
\font\tenssi=phvro

\textfont0=\tenss \scriptfont0=\sevenss \scriptscriptfont0=\fivess
\newfam\itfam \def\it{\fam\itfam\tenssi} 
\textfont\itfam=\tenssi
\def\dot{\mathaccent"70C7}
\newdimen\aadimen
\def\AA{\leavevmode\setbox0\hbox{h}\aadimen\ht0\advance\aadimen-1ex\setbox0
        \hbox{A}\rlap{\raise.67\aadimen\hbox to \wd0{\hss\char'27\hss}}A}
\font\text=phvr
\font\bold=phvb
\font\italic=phvro
\font\bigbold=phvb at 14.4pt
\font\bigtext=phvr at 12pt
\font\minibold=phvb at 8pt
\font\minitext=phvr at 8pt
%
\def\MAINTITLE#1{\footline={\hfill}\nopagenumbers\bigbold\parindent=0pt\hrule\vfill\par\advance\baselineskip by 7pt\par #1\par\advance\baselineskip by -7pt\par}

\def\AUTHOR#1{\vfill\bigtext\hrule\vfill#1}
\def\INSTITUTE#1{\vfill\parindent=1em\text\hrule\vfill#1}
\def\MAKETITLE#1{\vfill\parindent=0pt\hrule\vfill#1\vfill\hrule
    \eject\footline={\hfill\folio\hfill}\pageno=1}
\def\ABSTRACT#1{\pageno=1\parindent=20pt\bold\noindent Abstract.
    \text\vskip12pt\noindent#1\vfill\eject}
\def\TITLEA#1{\vskip24pt\bold\noindent #1\vskip12pt\text\noindent}

\def\BEGFIG#1{\midinsert\vskip#1}

\def\FIGURE#1#2{\par\advance\baselineskip by -2pt\noindent\minibold Figure
    #1.\ \minitext #2\par\advance\baselineskip by 2pt\par\text}
\def\SIDEFIGURE#1#2#3#4{\midinsert\par\advance\baselineskip by -2pt
    \par\hangindent=#2\hangafter=0\noindent\minibold Figure #3.\ \minitext
    #4\par\advance\baselineskip by 2pt\par\text\vskip#1}
\def\ENDFIG{\endinsert}

\def\TABCAP#1#2{\par\advance\baselineskip by -2pt\par\noindent\minibold Table
    #1.\ \minitext #2 \par\advance\baselineskip by 2pt\par\text\vskip2mm}

\def\BEGREF{\vfill\eject\TITLEA{References}}
\def\REF{\par\hangindent=20pt\parindent=0pt\hangafter=1}
\def\ENDREF{\parindent=20pt}
\def\la{\mathrel{\mathchoice {\vcenter{\offinterlineskip\halign{\hfil
    $\displaystyle##$\hfil\cr<\cr\noalign{\vskip1.5pt}\sim\cr}}}
    {\vcenter{\offinterlineskip\halign{\hfil$\textstyle##$\hfil\cr<\cr
    \noalign{\vskip1.0pt}\sim\cr}}}
    {\vcenter{\offinterlineskip\halign{\hfil$\scriptstyle##$\hfil\cr<\cr
    \noalign{\vskip0.5pt}\sim\cr}}}
    {\vcenter{\offinterlineskip\halign{\hfil$\scriptscriptstyle##$\hfil
    \cr<\cr\noalign{\vskip0.5pt}\sim\cr}}}}}

\def\dot{\mathaccent"70C7}
\MAINTITLE{Molecular Clouds Close to the Galactic Center}
\AUTHOR{Peter L.\ Biermann$^{\rm 1}$, Wolfgang J.\ Duschl$^{\rm 2,3}$,
Susanne von Linden$^{\rm 1,2}$}
\INSTITUTE{
\item{1:}Max-Planck-Institut f\"ur Radioastronomie, Auf dem H\"ugel 69, W-5300
Bonn, Germany
\item{2:}Institut f\"ur Theoretische Astrophysik, Universit\"at Heidelberg,
Im Neuenheimer Feld 561, W-6900 Heidelberg, Germany
\item{3:}Interdisziplin\"ares Zentrum f\"ur Wissenschaftliches Rechnen,
Universit\"at Heidelberg, Im Neuenheimer Feld 386, W-6900 Heidelberg, Germany}
\MAKETITLE{accepted for publication in \italic Astronomy and Astrophysics}
\ABSTRACT{We demonstrate that the accretion disk model for the Galactic Center
region by Linden et al (1993a) is applicable for at least one order of
magnitude in radius from the Galactic Center (10 \dots\ 100~pc). The
viscosity $\nu$ is shown to be weakly dependent on the radius $s$: $\nu \sim
s^{0.4}$. Finally, we discuss the influence of the inner boundary on the
structure of the inner disk regions.}
\TITLEA{Introduction}%
In a recent paper Linden et al. (1993a) (in the following often referred to as
LDB) have shown that one can use
molecular clouds in the Galactic Center (GC) region as tracers of the
large scale mass flow in this area. They demonstrated that their
technique allows one to determine the location of the clouds as well
as the radial mass flow rate  and the viscosity in the disk. While
they find large values for the viscosity, they can
show that these are in agreement with the observed velocity dispersion
and the scale height of the gas distribution.

Additionally, Linden et
al. (1993b) (=LBSLD) explained this large viscosity in terms of turbulence that
is induced through gravitational instability, subsequent star formation and
then actually driven by the ensuing supernova explosions. From the reasoning
given there it follows that one has to expect the viscosity $\nu$ to be smaller
for
smaller radii from the GC (for stationary accretion disks, $\nu\Sigma$ = const.
[see
sect.\ 4; $\Sigma$: vertically integrated surface density]; usually,
$\partial\Sigma
/\partial s < 0$ ($s$: radial coordinate), i.e.\ $\partial\nu /\partial s >
0$).

LDB demonstrated their method for one example, namely for one of the best
observed molecular clouds close to the GC, M-0.13-0.08. As it turned out
that this cloud is located in their model on a ring segment at an almost
constant radius of
$\approx$ 115~pc from the GC they could not draw any conclusions
about the radial variation of the viscosity.
In the present paper, we wish to test the validity of the LDB-model and the
parameter set as obtained by LDB by analyzing recent observations of
molecular clouds with small projected distances on either side of the
GC. The results for the H$_2$CO observations by Pauls et al. (1993)
demonstrate that these are in agreement with the model considerations.
In addition to determining the clouds'
geometrical positions to be $\approx$ 10~pc from the GC, this allows
even to get information about the gradient of the viscosity between
10 and 100~pc.

In the next section, we give a short description of the model as introduced
in LDB and the technique of their analysis that we also use here; in section
3, the observations and our results are
compiled. Section 4 is devoted to a discussion of these results, especially
of the implications for the radial variation of the viscosity. In the
subsequent section, we discuss how our conclusions depend on the physical
and numerical modelling of the disk's inner boundary. The last section,
finally, is devoted to a compilation of our conclusions and to an outlook
on how we plan to proceed in the analysis of the gas distribution in regions
close to the GC.
\TITLEA{The model for the gas dynamics close to the Galactic Center}%
Following LDB we model the gas and its distribution in the
innermost $\approx$ 200~pc of the Galaxy as a viscous accretion disk.
We assume that the disk extends along the galactic plane. We use the
following approximations as is usually done in the context of modelling
accretion disks (Weizs\"acker, 1943, L\"ust, 1952, Shakura and Sunyaev, 1973):
\item{$\bullet$}The disks are geometrically thin in the direction
perpendicular to the rotation plane. We take the standard approach and
describe the vertical structure in a one zone model.
\item{$\bullet$}The disk evolves in a gravitational potential that
is rotationally symmetric in the galactic plane. Despite the fact
that our Galaxy might be a barred spiral, the assumption of a rotationally
symmetric gravitational potential is reasonable as the length scales of such
a potential bar are larger than the ones of the disk that we are discussing
(Matsumoto et al., 1982; Blitz and Spergel, 1991; Binney et al., 1991).
Consequently, we assume the disk to be also rotationally symmetric. We
expect inhomogeneities in the disk flow to cause larger deviations from
rotaional symmetry than asymmetries in the galaxy's gravitational potential.
Nonetheless, in this first of analysis, we neglect such deviations from
axial symmetry.
\item{$\bullet$}We assume the radial mass flow rate,
$\dot M$, to be constant. Further, we assume that vertical infall
into the disk does not play a r\^ole at any radius.
These assumptions will be discussed in Sect.\ 5.

\noindent
In contrast to standard $\alpha$-accretion disks we allow for
several additional processes:
\item{$\bullet$} Viscosity is generated by processes outside the
accretion flow itself. In this respect it is a different type of accretion
disk model from the usually discussed $\alpha$-disks (Shakura and Sunyaev,
1973). Details of the viscosity prescription are discussed below.
\item{$\bullet$} If the disk becomes gravitationally unstable, i.e., if
Toomre's $Q$ value (Toomre 1964) drops below unity, we allow for star
formation to set in. We approximate this by dealing with a two phase
medium. The phase of the newly formed stars decouples from the viscous
evolution of the gaseous phase, and does not accrete, while it still
contributes to the potential.

\noindent
We describe the disk in a cylindrical coordinate system $\{s, z, \varphi\}$.

In the current model, we do not specify the
physical process that determines the
viscosity $\nu (s)$ in
a self-consistent way but rather choose some prescription for this
physical quantity.
This means that -- at this point -- we specify ``only'' the efficiency of
a process that causes radial transport of mass and angular momentum (i.e.,
that drives the accretion process).
In contrast to Linden
et al.\ (1993a), we do not require $\nu$ to be constant throughout the disk but
rather assume a power law dependence on the radius:

$$\nu = \nu_0 \left({s\over s_\nu}\right)^\beta\eqno{(1)}$$
$\nu_0$ and $s_\nu$ are scaling variables that have to be chosen such that
the resulting flow velocities are compatible with the results for
M-0.13-0.08. This requires

\BEGFIG{10.5cm}
\includegraphics{FIGURES/prefig1.ps}
\FIGURE{1}{Face-on view onto a cut out of the GC accretion disk. The observer
is located at $x$ = 8.5 kpc, $y$ = 0 pc. The (underlying) theoretical
radial velocity field is shown as a gray-scale picture (The definition of the
sign of the shown velocity component $v_x$ is given in the text). The two
vertical bars represent the observations as discussed in the text. An
individual observation consists of a direction, i.e., a vertical bar
in the figure at the proper position, and a radial velocity, i.e., a certain
shade of gray. In the framework of the model, the location of the cloud
components is determined by matching shades of gray.}
\ENDFIG

$$\nu_0 = 6\cdot 10^{26}\ {\rm cm^2 s^{-1}}\ \ \ \ \ \ \ \ {\rm for}\
\ \ \ \ \ \ \ s = s_\nu = 115\ {\rm pc}\eqno{(2)}$$
$\beta$ is the power that has to be determined through our analysis.
A meaningful determination of $\beta$ is only possible if the resulting
radial distance for the two gas components that we use for the analysis
(see next section) is considerably different from the 115~pc as determined
for M-0.13-0.08, as otherwise the radial baseline would be too short.

While, formally speaking, this prescripton for $\nu$ is introduced artificially
into LDB's accretion disk model, LBSLD have shown that one can
very well understand $\nu$ as to be caused by local processes in the region of
the disk that are due to the stellar component that is coexisting there
(star formation and supernovae stir the gas and thus enhance the turbulence
that acts as viscosity in the gas component).

In this paper, we assume that all other parameters are the same as used
by LDB:
\item{$\bullet$}There is a compact mass in the very GC with $M_{\rm c} =
2\cdot 10^6 M_\odot$, presumably a black hole.
\item{$\bullet$}The background mass distribution due to the stellar
component in the GC region is spherically symmetric and follows the power
law $M_{\rm b}(r) = 1.14\cdot 10^9 M_\odot \left( r / 100~{\rm pc}
\right)^{5/4}$.
$M_{\rm b}(r)$ is the mass enclosed in a sphere of radius $r$.
\item{$\bullet$}The disk's outer radius is located at 200~pc from the GC.
\item{$\bullet$}The disk's inner radius is located at 1~pc. In section
5, we shall discuss the influence of this parameter on our solutions, and
shall see that in the present context of the standard boundary condition,
the resulting positions of the gas clouds are far enough from the disk's
numerical boundary not to be influenced strongly by the boundary condition.
\item{$\bullet$} The disk is stationary with a radial mass flow rate of
$10^{24}$ g/s. Star formation does not consume an appreciable fraction of
the gas flowing in.

\noindent
In the framework of these assumptions and parameters, we solve for the
radial distribution of surface density and radial and azimuthal velocity
in the disk.

It turns out that Toomre's $Q$ value is never smaller than
unity. But we would like to remark that this is only a sufficient condition
for star formation actually to take place. As is shown in LBSLD (c.f.,
especially, their sect.\ 5), about
one and a half orders of magnitude lower a surface density might very well
be sufficient to sustain star formation under the conditions prevailing
in our disks. We shall discuss the consequences of a possible limit
cycle behaviour in a subsequent paper (Duschl et al., in preparation).

In principle, additional effects could come from non-stationary disks.
We have investigated this possibility and found that -- for our currently
still very limited sample of observed velocities -- this does not lead
to an improvement of our solutions. Actually, we have obtained all our
solutions by following the time evolution of disks from arbitrarily chosen
initial mass distributions until a steady state was reached. While we
followed the intermediate models in the course of the disk's evolution,
we found that stationary solutions usually accounted for the best
representations of the observations.

\TITLEA{Observations and results}%
To test the model of LDB and to refine the description further,
we attempt to include two (components of) gas clouds as
observed by Pauls et al. (1993) in H$_{\rm 2}$CO. Out of a large body
of observations in different frequencies and for different molecules (e.g.,
Bally et al., 1987 and 1988; Serabyn et al., 1987; Zylka et al., 1990;
Oda et al., private
communication), we choose this particular set observations mainly
on grounds of the following reasoning. To get meaningful information about the
radial gradient of the viscosity, we need a long radial baseline. As the cloud
M-0.13-0.08 is located at about 115 pc distance from the GC in a disk with
an outer radius of 200 pc, to achieve this long baseline, we need observations
of gas that is much closer to the GC than $\approx$ 100 pc. Moreover, this
choice allows to get information about the physical conditions in the
nearer vicinity of the GC than is obtained by working on M-0.13-0.08.

Before attempting to map large areas of the gas distribution in the GC,
such a step that confirms the large range applicability of the model
seems to us to be mandated.

The two components that we use have projected locations on either side of the
GC and both show radial velocities of -188.5 km/s (LSR velocities; negative
velocities pointing towards
us; velocities radial to the observer, but not to the GC); the one
component is located 24" west of the GC (in the following referred to
as component W-1), the other one 13" east of it (W-2).

We find that it is indeed possible to explain both observations in the
framework of the LDB model, and find that the two components are
located about 10~pc from the GC. The result is shown in Fig.\ 1. There
we followed the velocity definition of LDB, i.e., the velocities are
counted positive (in the reference system of an accretion disk centered
on the GC, a velocity
directed towards the GC from negative coordinates, i.e., from ``behind''
the GC has to be counted positive). Fig.\ 1 shows the plane of the disk
face-on. The observer at Earth is situated at $x = 8,500$ pc, and $y =
0$ pc. The vertical lines correspond to the observations which give us
a direction and a velocity. The velocity is coded by different shades
of grey. The underlying
theoretical velocity field as computed from the accretion disk model
is coded in the same way. The location of the clouds is determined
by matching grey shades. The shown velocities are $v_x$ velocities, i.e.,
LSR velocities radial with respect to the observer; this is meant by
the expression ``geocentric radial velocities''.

When keeping all the other parameters of the LDB model unchanged, we get
the best solutions for a radial dependence of the viscosity $\nu
\sim s^{0.4}$.

In the resulting models, the local galactocentric radial veliocities (i.e., the
real radial velocities in the disk with respect to the GC as opposed to the
above
defined geocentric ones that are radial with respect to the observer) are of
the
order of 10\%\ of the corresponding azimuthal (i.e., Keplerian) velocity. Only
for
radii smaller than $\approx$ 3~pc our approximation of a Keplerian,
geometrically
thin disk break down as then the radial velocities are no longer small compared
to the aziumthal ones. On the other side, in the radius range of $\approx$ 1~pc
our description becomes unsatisfactory anyway as there boundary effects will
play an important r\^ole. We address this aspect in Sect.\ 5.

\TITLEA{Discussion}%
Our results are in good agreement with what LDB found:
LDB determined a set of parameters by modelling the cloud M-0.13-0.08
which is located at $\approx$ 115 pc. We took the approach of assuming
that -- with the exception of the viscosity -- these parameters are
valid throughout the GC accretion disk. We demonstrated that it is indeed
possible to understand the observed velocities of the H$_{\rm 2}$CO clouds
at $\approx$ 10 pc from the GC.

On theoretical grounds (e.g., LBSLD) we expected a radial variation of $\nu
\sim s^\beta$. The exponent $\beta$ itself is determined through our present
analysis.
It turns out that for all $\beta \in [0, 1]$ the velocity of the western
component (W-1) can easily be reproduced in the framework of the LDB-model.
But only for $\beta \la 0.45$, we succeed to get good representation for
both components. Moreover, we find that for $\beta \la 0.3$, the solutions
become meaningless as
then the radial velocities at the points of the clouds are no longer
small compared to the azimuthal velocities. Then the approximations of the
model calculations break down. While on the first glance this is mainly
a problem of the model's assumptions and does not exclude a predominantly
radial motion of the gas, one has to keep in mind that observations clearly
indicate that a disk-like structure of the gas is maintained down to the radius
of the molecular ring at 2~pc. This indicates that a value for $\beta \approx
0.4$ is a physically consistent and reasonable result.

This result is also in reasonable agreement with the results of LBSLD.
There it was concluded on the grounds of order of magnitude estimates
that $\nu$ should vary approximately linearly with the radius. Our results
indicate that the radial dependence is somewhat weaker.

This can be understood if we loosen LBSLD's assumption of an inversely linear
radial dependence of $\Sigma$ but rather take a more general approach.
The continuity equation $\dot M =$ $2 \pi s \Sigma v_s$, together with
the approximation $v_s \approx$ $s / \tau_{\rm visc} \approx$ $\nu / s$
gives for stationary accretion disks $\nu \Sigma = $ const. ($v_s$: radial
velocity of matter in the accretion disk; $\tau_{\rm visc}$ viscosity
time scale charaterizing the typical time scale for matter motion in
the accretion disk vom $s$ to the disk's inner edge). Additionally,
$\nu \sim h c_s$. If we now again follow LBSLD, we get a selfconsistent
solution with $c_s \sim$ $s^{-0.17}$, $v_\varphi \sim $ $s^{1/8}$, and $h
\sim $ $s^{1.04}$, and
find that $\nu \sim s^{0.54}$ and $\Sigma \sim$ $s^{-0.54}$ which is in
good
agreement with the value for $\beta$ that we deduced from observations
($c_s$: (isothermal) sound velocity; $v_\varphi$: azimuthal (Keplerian)
velocity in the disk; $h$: vertical (half-)thickness of the disk). It is
important to note that the above analytical derivation for $\beta$ and
the numerical determination are fully independent of each other. Thus
this is not only a pleasing matching of numbers but a very strong indication
that the physical mechanism of supernova driven turbulence as discussed by
LBSLD indeed can reproduce what is going on in the GC region.

\TITLEA{Physics and numerics of the disk's inner boundary}%
We have placed the (numerical) inner radius of the disk at 1~pc. There we
applied the standard accretion disk boundary condition that matter leaving
the Keplerian disk in inward direction keeps all its angular momentum and
thus takes it away from the disk. This also means that this matter keeps
its orbital kinetic energy.

This approximation is known to be numerically a problematic one (as it
leads to a singularity at the boundary itself) and physically an
oversimplification (Duschl and Tscharnuter, 1991). On the other hand
it is known that it leads to serious problems only at radii close to
the inner radius (i.e, in our case, for $s \not\gg 10$ pc). To ensure
that our results are not dominated by the numerical boundary condition,
we repeated our calculations for an inner radius of the disk of 0.01~pc.
The results were only slightly influenced at the radii where the two
clouds are located.

While this is a sufficient numerical justification in the context of the
questions discussed in this paper, this kind of a physically and numerically
doubtful boundary condition is a conceptual problem for the entire description.
We shall address this problem in a forthcoming paper (Duschl et al., in
preparation). There, from the formal point of view, we will take the
approach of assuming that only a fraction $\zeta \in [0, 1]$ of the angular
momentum of the innermost orbit is transported away. From the physical point
of view, we will demonstrate that under the conditions prevailing in this
radial range from the GC, it is even very likely that the matter will leave
as much angular momentum as possible in the disk. The maximum will be
mainly determined by the fact that -- in addition to conservation of angular
momentum -- also energy conservation influences the global disk structure.
For the present context, it is important to note that the location
of the disk's numerical inner boundary at 1~pc does not influence the
results very much.

In the present paper, we follow LDB and assume that the mass flow rate
reaching the disk radially in the rotational plane ($\dot M$) does not
vary with time,  and that the vertical mass infall into the disk
is negligible at all radii (cf.\ Sect.\ 2). Within the framework of our
approximations, the latter
means that the integral of vertical mass infall over the relevant disk
surface area is always considerably smaller than the radial mass flow
$\dot M$. We assumed $\dot M$ to be a constant as, in the present context,
we were mainly interested in investigating the radial variations of the
viscosity. In a later phase of our analysis of the gas dynamics in the
vicinity of the GC region, this approximation has to be loosened. This
then could lead to non-stationary accretion disk models.

\TITLEA{Conclusions}%
We have demonstrated that the accretion disk model for the gas flow in the GC
region by
LDB is valid over at least an order of magnitude in radius from the GC
if one allows for a (weak) radial variation of the viscosity. We also
show that the deduced radial variation of the effective turbulent viscosity
is compatible with the scenario of a supernova driven turbulence as proposed
by LBSLD.

These results make
it worthwile to attempt to construct a map of the distribution of molecular
gas in the innermost $\approx$ 200~pc around the GC. For radii of the
order of 10~pc or even smaller, the question of a proper (numerical) inner
boundary condition becomes of importance. The standard boundary condition
of accretion disk theory is -- as in almost all other cases where accretion
disks play a role -- not very well suited. In the case of the GC the existence
of a (comparatively sharp) inner edge of the disk at $\approx$ 2~pc gives an
important constraint.
\TITLEA{Acknowledgements}%
We benefitted very much from discussions with Drs.\ R.\ Wielebinski
and T.L.\ Wilson. WJD acknowledges financal support during a stay at the
MPIfR. This work was in part supported by the Deutsche Forschungsgemeinschaft
through {\it Sonderforschungsbereich 328} ``Evolution of galaxies'' (SvL, WJD)
and by the Bundesministerium f\"ur Forschung und Technologie through a
{\it Verbundforschungs-}grant (PH/055HD45A, WJD).
\BEGREF
\REF Bally J., Stark A.A., Wilson R.W., and Henkel C., 1987, ApJSuppl 65, 13
\REF Bally J., Stark A.A., Wilson R.W., and Henkel C., 1988, ApJ 324, 223
\REF Binney J., Gerhard O.E., Stark A.A., Bally J., and Uchida K.I., 1991,
MNRAS 252, 210
\REF Blitz L., Spergel D.N., 1991, ApJ 379, 631
\REF Duschl W.J., Tscharnuter W.M., 1991, A\&A 241, 153
\REF Linden S.v., Duschl W.J., Biermann P.L., 1993a, A\&A 269, 169 (=LDB)
\REF Linden S.v., Biermann P.L., Schmutzler T., Lesch H., Duschl W.J.,
1993b, A\&A submitted (=LBSLD)
\REF L\"ust R., 1952, Zeitschr.\ f.\ Naturforschung 7a, 87
\REF Matsumoto T., Hayakawa S., Koizumi H., Murakami H., Uyama K., Yamagami T,
and Thomas J.A., 1982, in: The Galactic Center (Eds.: Riegler G.R., Blandford
R.D.), 48
\REF Pauls T., Johnston K.J., Wilson T.L., Marr J.M., Rudolph A., 1993,
ApJ 403, L13
\REF Serabyn E., G\"usten R., Walmsley C.M., Wink J.E., and Zylka R., 1987,
A\&A 169, 85
\REF Shakura N.I., Sunyaev R.A., 1973, A\&A 24, 337
\REF Toomre A., 1964, ApJ 139, 1217
\REF Weizs\"acker C.F.\ von, 1943, Zeitschr.\ f.\ Astrophys.\ 22, 319
\REF Zylka R., Mezger P.G., and Wink J.E., 1990, A\&A 234, 133
\ENDREF
\bye